\RequirePackage{lineno}
\documentclass[aps,twocolumn,showpacs,byrevtex,prl,reprint]{revtex4-1}

\usepackage{graphicx}
\usepackage{dcolumn}
\usepackage{bm}
\usepackage{rotating}
\usepackage{epstopdf}
\usepackage{color}
\usepackage{verbatim} 
\usepackage{multirow}
\usepackage[abs]{overpic}
\usepackage{amsmath}
\usepackage{amssymb}
\usepackage{xspace}
\usepackage[titletoc]{appendix}
\usepackage{float}

\usepackage[colorlinks,
            linkcolor=blue,
            anchorcolor=blue,
            citecolor=blue]{hyperref}

\newcommand{\PreserveBackslash}[1]{\let\temp=\\#1\let\\=\temp}
\newcolumntype{C}[1]{>{\PreserveBackslash\centering}p{#1}}
\newcolumntype{R}[1]{>{\PreserveBackslash\raggedleft}p{#1}}
\newcolumntype{L}[1]{>{\PreserveBackslash\raggedright}p{#1}}

\newcommand{\EE}{e^+e^-}

\newcommand{\too}{\rightarrow}

\newcommand{\BESIIIorcid}[1]{\href{https://orcid.org/#1}{\hspace*{0.1em}\raisebox{-0.45ex}{\includegraphics[width=1em]{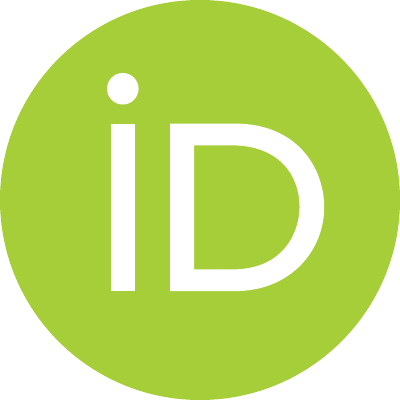}}}}

\begin{document}
\graphicspath{{figure/}}
\DeclareGraphicsExtensions{.eps,.png,.ps}
\title{\boldmath Study of $\bar{\Lambda}$-$p$ Annihilation into Light Mesons}
\author{
  \begin{small}
    \begin{center}
      M.~Ablikim$^{1}$\BESIIIorcid{0000-0002-3935-619X},
M.~N.~Achasov$^{4,d}$\BESIIIorcid{0000-0002-9400-8622},
P.~Adlarson$^{81}$\BESIIIorcid{0000-0001-6280-3851},
X.~C.~Ai$^{87}$\BESIIIorcid{0000-0003-3856-2415},
C.~S.~Akondi$^{31A,31B}$\BESIIIorcid{0000-0001-6303-5217},
R.~Aliberti$^{39}$\BESIIIorcid{0000-0003-3500-4012},
A.~Amoroso$^{80A,80C}$\BESIIIorcid{0000-0002-3095-8610},
Q.~An$^{77,64,\dagger}$,
Y.~H.~An$^{87}$\BESIIIorcid{0009-0008-3419-0849},
Y.~Bai$^{62}$\BESIIIorcid{0000-0001-6593-5665}
% ...
\\
\vspace{0.2cm}
(BESIII Collaboration)\\
\vspace{0.2cm} {\it
$^{1}$ Institute of High Energy Physics, Beijing 100049, People's Republic of China\\
% ...
$^{4}$ Budker Institute of Nuclear Physics SB RAS (BINP), Novosibirsk 630090, Russia\\
% ...
$^{31}$ INFN Sezione di Ferrara, (A)INFN Sezione di Ferrara, I-44122, Ferrara, Italy; (B)University of Ferrara, I-44122, Ferrara, Italy\\
% ...
$^{39}$ Johannes Gutenberg University of Mainz, Johann-Joachim-Becher-Weg 45, D-55099 Mainz, Germany\\
% ...
$^{62}$ Southeast University, Nanjing 211100, People's Republic of China\\
% ...
$^{64}$ State Key Laboratory of Particle Detection and Electronics, Beijing 100049, Hefei 230026, People's Republic of China\\
% ...
$^{77}$ University of Science and Technology of China, Hefei 230026, People's Republic of China\\
% ...
$^{80}$ University of Turin and INFN, (A)University of Turin, I-10125, Turin, Italy; (B)University of Eastern Piedmont, I-15121, Alessandria, Italy; (C)INFN, I-10125, Turin, Italy\\
$^{81}$ Uppsala University, Box 516, SE-75120 Uppsala, Sweden\\
% ...
$^{87}$ Zhengzhou University, Zhengzhou 450001, People's Republic of China\\
\vspace{0.2cm}
% ...
$^{d}$ Also at the Novosibirsk State University, Novosibirsk, 630090, Russia\\
% ...
      }\end{center}
    \vspace{0.4cm}
\end{small}
}
\affiliation{}


\begin{abstract}
The annihilation of $\bar{\Lambda}$-$p$ into light mesons is studied using $(10.087\pm0.044)\times10^{9}$ $J/\psi$ events collected with the BESIII detector at the BEPCII storage ring. The $\bar{\Lambda}$ baryons are produced in the process $J/\psi\too\Lambda\bar{\Lambda}$ and interact with the protons in the hydrogen nuclei in the cooling oil of the beam pipe. The reactions $\bar{\Lambda}p\too K^+\pi^+\pi^-$ and $K^+2\pi^+2\pi^-$ are observed for the first time, evidence for the reaction $\bar{\Lambda}p\too K^0_S2\pi^+\pi^-$ is found, and no significant signals are seen for the reactions $\bar{\Lambda}p\too K^0_S\pi^+$, $K^+3\pi^+3\pi^-$ and $2K^+K^-$. The cross sections of these reactions are also measured at $P_{\bar{\Lambda}}\approx 1.074$~GeV/$c$. Furthermore, no significant structures are observed in the combinations of the final state particles for these reactions. The annihilation of $\bar{\Lambda}$-$p$ into light mesons presents the similar behavior as $\bar{n}$-$p$ reactions in this energy region. These new measurements will serve as important inputs for the theoretical understanding of the antihyperon-nucleon interaction, and further contribute to understanding antibaryon-baryon interaction in a unified perspective.
\end{abstract}

\maketitle

The measurement of hyperon-nucleon scattering is crucial to understand quantum chromodynamics, comprehend the nature of hypernuclei~\cite{hypernuclei1, hypernuclei2, hypernuclei3, hypernuclei4}, and solve the so-called ``hyperon puzzle" concerning the masses of neutron stars~\cite{neutronstar1, neutronstar2, neutronstar3, neutronstar4, neutronstar5}. Hyperon-nucleon interaction has also been studied in different theoretical models, such as the constituent quark model~\cite{theory1, theory2, theory3}, the meson-exchange picture~\cite{theory4, theory5, theory6}, and the chiral effective field theory approach~\cite{theory7, theory8, theory9, theory10, theory11, theory12, theory13}. However, progress in this field is hindered by the limited availability and short lifetime of hyperon beams, resulting in slow advancements~\cite{pdg}. There is a large uncertainty in the understanding of hyperon-nucleon interaction due to the lack of relevant scattering data. Hence experimental measurements in this area are a critical input to constrain the calculations of hyperon-nucleon interactions, and drive the research in this arena further.

Compared to hyperon-nucleon scattering, the situation is even worse for antihyperon-nucleon scattering due to the absence of antihyperon sources. Recently, BESIII completed the first study of antihyperon-nucleon scattering, using the reaction $\bar{\Lambda}p\too\bar{\Lambda}p$ and found a strong forward peak for the reaction at $P_{\bar{\Lambda}}\approx 1.074$ GeV/$c$~\cite{lambdabarscatteringbes}. This new result has stimulated some discussions about antihyperon-nucleon interaction in theory~\cite{theory14, theory15, theory16}. Therefore, more experimental data will improve the understanding of the antihyperon-nucleon interaction, and can provide complementary constraints on the hyperon-nucleon interaction theory. Many $\bar{n}$-$p$ reactions have been measured in experiments~\cite{nbarp1, nbarp2, nbarp3, nbarp4, nbarp5}, especially for $\bar{n}$-$p$ annihilation into multiple $\pi$ meson final states. A natural extension is to measure $\bar{\Lambda}$-$p$ annihilation into light mesons. The difference between them is that one $\bar{d}$ quark in the antineutron is replaced with $\bar{s}$, while the charge remains unchanged. If the $\bar{s}$ in $\bar{\Lambda}$ and the corresponding $\bar{d}$ in antineutron are just bystanders in the reactions, the behaviors of $\bar{\Lambda}$-$p$ and $\bar{n}$-$p$ annihilations should be similar. So it is also interesting to compare the results of $\bar{\Lambda}$-$p$ and $\bar{n}$-$p$ reactions together, which is very useful to understand antibaryon-baryon interaction in a unified perspective. Additionally, we also search for interesting excited states in the combinations of the final state mesons for $\bar{\Lambda}$-$p$ reactions.

In this Letter, we present a study of the reactions $\bar{\Lambda}p\too K^0_S\pi^+$, $K^+\pi^+\pi^-$, $K^0_S2\pi^+\pi^-$, $K^+2\pi^+2\pi^-$, $K^+3\pi^+3\pi^-$, and $2K^+K^-$. The cross sections of these reactions are all measured for the first time, and compared with the results of $\bar{n}$-$p$ reactions.

The BESIII detector records symmetric $e^+e^-$ collisions at the BEPCII collider~\cite{bepcii}. Details of the BESIII detector can be found in Ref.~\cite{besiii}. Intense almost monoenergetic $\bar{\Lambda}$ hyperons with a momentum of $1.074$~GeV/$c$ within a range of $\pm0.017$~GeV/$c$ can be produced via the decay $J/\psi\too\Lambda\bar{\Lambda}$ using the $(10.087\pm0.044)\times10^{9}$ $J/\psi$ events collected at BESIII~\cite{totalnumber}. The $\bar{\Lambda}$ baryons can interact with the material in the beam pipe, which is composed of gold ($^{197}\rm{Au}$), beryllium ($^{9}\rm{Be}$) and oil (number of atoms $^{12}\rm{C}:$$^{1}\rm{H}$$=1:2.13$)~\cite{lambdabarscatteringbes}. Based on this novel method~\cite{lear, method1, method2}, some (anti)hyperon-nucleon reactions have been conducted at BESIII~\cite{xiscatteringbes, lambdabarscatteringbes, lambdascatteringbes, sigmascatteringbes}. In particular, utilizing the almost static protons in the $^{1}\rm{H}$ of the cooling oil of the beam pipe, the information on the $\bar{\Lambda}$-$p$ reactions can be directly extracted~\cite{lambdabarscatteringbes}.

In this analysis, simulated data samples are produced with a {\sc{Geant4}}-based~\cite{geant4} Monte Carlo (MC) package, which includes the geometric description of the BESIII detector~\cite{besiii} and the detector response. They are used to determine detection efficiencies and to estimate backgrounds. The simulation models the beam energy spread and initial state radiation in the $e^+e^-$ annihilations with the generator {\sc kkmc}~\cite{KKMC}. The inclusive MC sample includes both the production of the $J/\psi$ resonance and the continuum processes incorporated in {\sc kkmc}~\cite{KKMC}. All particle decays are modeled with {\sc evtgen}~\cite{ref:evtgen} using branching fractions either taken from the Particle Data Group (PDG)~\cite{pdg}, where available, or otherwise estimated with {\sc lundcharm}~\cite{ref:lundcharm}. Final state radiation from charged final state particles is incorporated using the {\sc photos} package~\cite{photos}. The signal process considered in this analysis is $J/\psi\too\Lambda\bar{\Lambda}$ with $\Lambda\too p\pi^-$ and $\bar{\Lambda}p\too K^0_S\pi^+$, $K^+\pi^+\pi^-$, $K^0_S2\pi^+\pi^-$, $K^+2\pi^+2\pi^-$, $K^+3\pi^+3\pi^-$, $2K^+K^-$, where $K^0_S$ is reconstructed via the decay $K^0_S\too\pi^+\pi^-$. In the signal simulation, the angular distribution of $J/\psi\too\Lambda\bar{\Lambda}$ is generated according to the measurement in Ref.~\cite{alpha}. We simulate the reactions $\bar{\Lambda}p\too$ light mesons by taking the proton to be at rest, and using a phase-space distribution.

Charged tracks detected in the multilayer drift chamber (MDC) are required to be within a polar angle ($\theta$) range of $|\cos\theta|<0.93$, where $\theta$ is the angle between the charged track and the $z$-axis, which is the symmetry axis of the MDC. Particle identification (PID) for charged tracks combines measurements of the energy loss (d$E$/d$x$) in the MDC and the flight time in the time-of-flight system (TOF) to form likelihoods $\mathcal{L}(h)$ $(h=p, K, \pi)$ for each hadron $h$ hypothesis. Tracks are identified as protons when the proton hypothesis has the largest likelihood $(\mathcal{L}(p)>\mathcal{L}(K)$ and $\mathcal{L}(p)>\mathcal{L}(\pi))$, while charged kaons and pions are identified by comparing the likelihoods for the kaon and pion hypotheses, $\mathcal{L}(K)>\mathcal{L}(\pi)$ and $\mathcal{L}(\pi)>\mathcal{L}(K)$, respectively.

The total number of good charged tracks for candidate events is required to be equal to the number of charged particles in the reconstructed final state of each signal process. We apply the PID requirements on these charged tracks, and the particle type is required to be consistent with the final state of each signal process. For the decay $\Lambda\too p\pi^-$, we perform the vertex fit of $\Lambda$ by considering all $p\pi^-$ combinations. The $p\pi^-$ combination with the smallest value of $|M(p\pi^-)-m_{\Lambda}|$, where $m_{\Lambda}$ is the $\Lambda$ nominal mass, is taken as $\Lambda$ candidate. In this Letter, all nominal masses are taken from the PDG~\cite{pdg}. The $\Lambda$ signal region is defined as $|M(p\pi^-)-m_{\Lambda}|<0.003$~GeV/$c^{2}$, and the $\Lambda$ sideband region is defined as $0.006<|M(p\pi^-)-m_{\Lambda}|<0.009$~GeV/$c^{2}$. For the reactions $\bar{\Lambda} p\too$ light mesons where there is no $K^0_S$ meson in the final state, we perform the vertex fit for all remaining charged particles. If there is a $K^0_S$ meson in the final state, we first perform the vertex fit of $K^0_S$ by considering all $\pi^+\pi^-$ combinations, and then perform the second vertex fit for $K^0_S$ mesons and all remaining charged particles. The $\pi^+\pi^-$ track combination with the minimum value of $|M(\pi^+\pi^-)-m_{K^0_S}|$, where $m_{K^0_S}$ is the $K^0_S$ nominal mass is taken as the $K_S^0$ candidate. The $K^0_S$ signal region is defined as $|M(\pi^+\pi^-)-m_{K^0_S}|<0.015$~GeV/$c^{2}$, and the $K^0_S$ sideband region is defined as $0.030<|M(\pi^+\pi^-)-m_{K^0_S}|<0.045$~GeV/$c^{2}$.

To select the signal events of $J/\psi\too\Lambda\bar{\Lambda}$, the invariant mass recoiling against the $\Lambda$, $M_{\text{recoil}}(\Lambda)$, is required to be in the $\bar{\Lambda}$ signal region, defined as \mbox{$|M_{\text{recoil}}(\Lambda)-m_{\bar{\Lambda}}|<0.020$~GeV/$c^{2}$}, where \mbox{$M_{\text{recoil}}(\Lambda)\equiv\sqrt{E^2_{\text{beam}}-|\vec{P}_{\Lambda}c|^2}/c^2$}, $E_{\text{beam}}$ is the $e^{\pm}$ beam energy, and $\vec{P}_{\Lambda}$ is the measured momentum of the $\Lambda$ candidate in the $\EE$ rest frame. The $M_{\text{recoil}}(\Lambda)$ sideband regions are defined as $0.040<|M_{\text{recoil}}(\Lambda)-m_{\bar{\Lambda}}|<0.060$~GeV/$c^{2}$. To select those signal events that react with the cooling oil in the beam pipe, the $R_{xy}$ signal region is defined as $[3.0, 3.5]$~cm~\cite{lambdabarscatteringbes}, taking into account the detector resolution, where $R_{xy}$ is the distance from the reconstructed reaction vertex to the $z$-axis. To remove the events from the reactions between $\bar{\Lambda}$ and $^{197}\rm{Au}/^{9}\rm{Be}/^{12}\rm{C}$ nuclei, we define the momentum of the proton in the $^{1}\rm{H}$ of the cooling oil as $P(p_{\rm{oil}})\equiv|\vec{P}_{i}-(\vec{P}_{e^+e^-}-\vec{P}_{\Lambda})|$, where $\vec{P}_{i}$ represents the total momentum of the light mesons in each reaction and $\vec{P}_{e^+e^-}$ is the total momentum of the $e^+e^-$ beams. As the proton in the $^{1}\rm{H}$ of the cooling oil is effectively static, while protons in $^{197}\rm{Au}/^{9}\rm{Be}/^{12}\rm{C}$ nuclei have Fermi momentum, $P(p_{\rm{oil}})$ is expected to be around zero for signal processes but hundreds of MeV/$c$ for background processes. To remove these events, the requirement $P(p_{\rm{oil}})<0.04$~GeV/$c$ is applied.

\begin{figure*}[htbp]
\begin{center}
\begin{overpic}[width=0.92\textwidth]{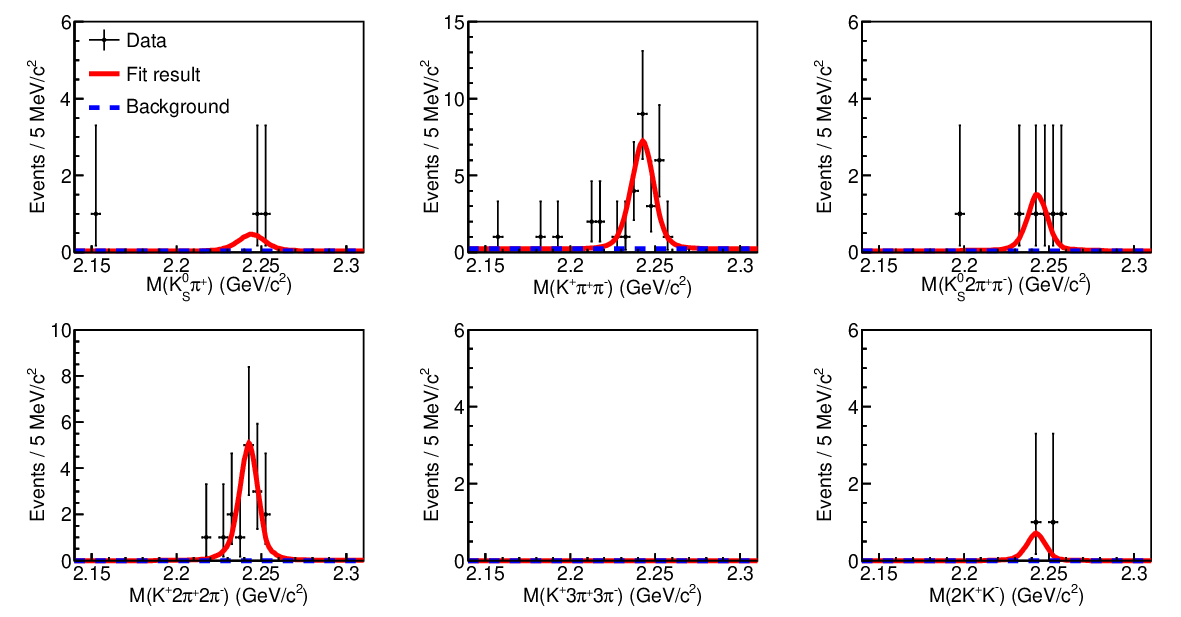}
\end{overpic}
\caption{Invariant mass distributions of the light mesons for each reaction in data (denoted by dots with error bars). The solid red curves are the total fit results and the dashed blue curves are the background components.}
\label{fig:fit_result}
\end{center}
\end{figure*}

The momentum of the incident $\bar{\Lambda}$ hyperons from the decay of the $J/\psi$ meson is around $1.074$~GeV/$c$. Hence the center-of-mass energies for the incident $\bar{\Lambda}$ and a static proton are all $2.243$~GeV/$c^2$ within a range of $\pm0.005$~GeV/$c^2$. Figure~\ref{fig:fit_result} shows the invariant mass distributions of the light mesons in each reaction from data after the final event selection. Clear signals for the reactions $\bar{\Lambda}p\too K^+\pi^+\pi^-$ and $K^+2\pi^+2\pi^-$ are seen around $2.243$~GeV/$c^2$, and there are some signal events for the reaction $\bar{\Lambda}p\too K^0_S2\pi^+\pi^-$, while no significant signals are found for the reactions $\bar{\Lambda}p\too K^0_S\pi^+$, $K^+3\pi^+3\pi^-$ and $2K^+K^-$. There are almost no background events in the $J/\psi$ inclusive MC sample and no events remaining in the data sidebands for the $M(p\pi^-)$, $M(\pi^+\pi^-)$, and $M_{\text{recoil}}(\Lambda)$ variables. To determine the signal yield, unbinned maximum likelihood fits are performed to the invariant mass distributions. We use the MC-determined shape to describe the signal, where the yield acts as a free fit parameter. The background is described by a uniform distribution with the number of events as free parameter. The fit results are shown in Fig.~\ref{fig:fit_result}, and are listed in Table~\ref{tab:result}. The significances listed in Table~\ref{tab:result} have already taken the systematic uncertainties into account. For the reactions with significances less than $3\sigma$, we also give the upper limits at $90\%$ confidence level (C.L.) using a Bayesian method. With the fit function described before, we scan the signal yield to obtain the likelihood distribution, and smear it with the systematic uncertainty. The upper limit of the signal yield $N^{\rm{ul}}$ at $90\%$ C.L. is obtained via $\int_{0}^{N^{\rm{ul}}}F(x){\rm{d}}x/\int_{0}^{\infty}F(x){\rm{d}}x=0.90$, where $F(x)$ is the probability density function of the likelihood distribution. Furthermore, we also check the invariant mass distributions of all final state particle combinations for the reactions $\bar{\Lambda}p\too K^+\pi^+\pi^-$ and $K^+2\pi^+2\pi^-$, but no significant structures are observed, and the data is consistent with the signal MC based on the current statistics.

\begin{table}[htbp]
\renewcommand{\arraystretch}{1.4}
\begin{center}
\caption{Number of signal events $N$ (or upper limits at 90\% C.L.), efficiencies $\epsilon$, significances $S$ for the measured $\bar{\Lambda} p\too$ light mesons. The uncertainties of $N$ are statistical only.}
\label{tab:result}
\begin{tabular}{llll}
  \hline
  \hline
  $\bar{\Lambda}p\too$ & $N$ & $\epsilon$ (\%) & $S$ ($\sigma$) \\
  \hline
  $K^{0}_{S}\pi^{+}$           & $1.8^{+1.8}_{-1.1}(<5.2)$  & 23.6 & $1.9$ \\
  $K^{+}\pi^{+}\pi^{-}$        & $24.3\pm5.3$               & 18.7 & $7.4$ \\
  $K^{0}_{S}2\pi^{+}\pi^{-}$   & $4.6^{+2.7}_{-2.1}$        & 7.3  & $3.0$ \\
  $K^{+}2\pi^{+}2\pi^{-}$      & $15.0\pm3.9$               & 12.1 & $6.6$ \\
  $K^{+}3\pi^{+}3\pi^{-}$  \ \ \    & $0.0^{+1.1}_{-0.0}(<2.3)$ \ \ \  & 3.9 \ \ \ \ \ & $0.0$ \\
  $2K^{+}K^{-}$                & $2.0^{+1.8}_{-1.1}(<5.4)$  & 18.2 & $2.3$ \\
  \hline
  \hline
\end{tabular}
\end{center}
\end{table}

Using the same method as in Ref.~\cite{lambdabarscatteringbes}, the cross sections, $\sigma$, of the reactions $\bar{\Lambda}p\too$ light mesons can be calculated with
\begin{linenomath*}
\begin{equation}
    \sigma = \frac{N}{\epsilon \mathcal{B}(\Lambda\too p\pi^-) \mathcal{B}_{K} \mathcal{L}_{\rm{eff}} },
\end{equation}
\end{linenomath*}
where $\epsilon$ is the selection efficiency provided in Table~\ref{tab:result}, $\mathcal{B}_{K}=\mathcal{B}(K^0_S\too \pi^+\pi^-)$ if there is $K^0_S$ meson in the final state, otherwise $\mathcal{B}_{K}=1$, and $\mathcal{L}_{\rm{eff}}$ is the effective luminosity of the reaction of the $\bar{\Lambda}$ flux produced from $J/\psi\too\Lambda\bar{\Lambda}$ with the target material:
\begin{linenomath*}
\begin{equation}
    \mathcal{L}_{\rm{eff}} = \frac{\it{N}_{\it{J}/\psi}\mathcal{B}_{\it{J}/\psi}}{\rm{2}+\frac{\rm{2}}{\rm{3}}\alpha} \int_{a}^{b}\int_{\rm{0}}^{\pi} (\rm{1}+\alpha \rm{cos}^2\theta) \it{e}^{-\frac{x}{\rm{sin}\theta \it{\beta\gamma L}}}N_{\rm{H}}\rm{d}\theta \rm{d}\it{x}.
\end{equation}
\end{linenomath*}
In the integral of this formula, the angular distribution of the $\bar{\Lambda}$ flux, the attenuation of the $\bar{\Lambda}$ flux, and the number of target nuclei are considered in turn. The number of $J/\psi$ events is denoted $N_{J/\psi}$, $\mathcal{B}_{J/\psi}$ is the branching fraction of $J/\psi\too\Lambda\bar{\Lambda}$, $\alpha$ is the parameter of the angular distribution of $J/\psi\too\Lambda\bar{\Lambda}$~\cite{alpha}, $\beta\gamma\equiv \frac{\sqrt{E_{\rm{beam}}^{\rm{2}}-m_{\rm{\bar{\Lambda}}}^{\rm{2}}c^{\rm{4}}}}{m_{\rm{\bar{\Lambda}}}c^{\rm{2}}}$ is the ratio of the momentum to the mass of the $\bar{\Lambda}$, and $L\equiv c\tau$ is the product of the speed of light and the mean lifetime of the $\bar{\Lambda}$~\cite{pdg}. The number of target nuclei per unit volume is $N_{\rm{H}}$, $a$ and $b$ are the distances from the inner surface and outer surface of the cooling oil in the beam pipe to the $z$-axis, and $\theta$ and $x$ are the angle and distance to the $z$-axis~\cite{lambdabarscatteringbes}. The beam pipe can be regarded as infinitely long with respect to the product of $\beta\gamma L$ for $\bar{\Lambda}$. These parameters can be found in Ref.~\cite{lambdabarscatteringbes}.

The measured cross sections of $\bar{\Lambda} p\too$ light mesons are listed in Table~\ref{tab:result2}. For convenient comparison, we also include the measured cross sections of $\bar{n} p\too$ light mesons in $0.5\leq P_{\bar{n}}\leq 0.8$~GeV/$c$ from Ref.~\cite{nbarp1}. Based on the current measurements of $\bar{n} p$ reactions, these results are the most suitable for comparison.
A similar trend is found in this energy region for the five reactions $\bar{\Lambda}p\too KN\pi$ and seven reactions $\bar{n}p\too \pi N\pi$ ($N\geq1$). Firstly, the cross sections increase rapidly from $N=1$ to $N=2$, then become comparable or relatively stable as $N$ increases; finally, the cross sections decreases rapidly as $N$ further increases, the critical number with a significant decrease in cross sections is about $N=6$. This similar behavior may imply the antistrange quark in these $\bar{\Lambda}$-$p$ annihilations is just a bystander, and does not directly participate in annihilation. These results should be useful to understand antibaryon-baryon interaction in a unified perspective.

\begin{table}[htbp]
\renewcommand{\arraystretch}{1.4}
\begin{center}
\caption{Cross sections (or upper limits at 90\% C.L.) for the measured $\bar{\Lambda} p\too$ light mesons in this work, together with the results for $\bar{n} p\too$ light mesons from Ref.~\cite{nbarp1}. The first uncertainties are statistical, and the second systematic.}
\label{tab:result2}
\begin{tabular}{cccc}
  \hline
  \hline
  \multicolumn{2}{c}{$P_{\bar{\Lambda}}\approx 1.074$~GeV/$c$} & \multicolumn{2}{c}{$0.5\leq P_{\bar{n}}\leq 0.8$~GeV/$c$} \\
  $\bar{\Lambda}p\too$ & $\sigma$ (mb) & $\bar{n}p\too$ & $\sigma$ (mb) \\ \hline
  $K^{0}_{S}\pi^{+}$         & $0.22^{+0.22}_{-0.13}\pm0.02(<0.62)$ & $\pi^{+}\pi^{0}$          & $0.16\pm0.05$ \\
  $K^{+}\pi^{+}\pi^{-}$      & $2.54\pm0.56\pm0.23$                 & $2\pi^{+}\pi^{-}$         & $1.16\pm0.10$ \\
  $K^{0}_{S}2\pi^{+}\pi^{-}$ & $1.78^{+1.05}_{-0.81}\pm0.20$        & $2\pi^{+}\pi^{-}\pi^{0}$  & $10.15\pm0.51$ \\
  $K^{+}2\pi^{+}2\pi^{-}$    & $2.43\pm0.63\pm0.27$                 & $3\pi^{+}2\pi^{-}$        & $2.41\pm0.18$ \\
  $K^{+}3\pi^{+}3\pi^{-}$    & $0.00^{+0.55}_{-0.00}\pm0.00(<1.16)$ & $3\pi^{+}2\pi^{-}\pi^{0}$ & $4.86\pm0.31$ \\
  $2K^{+}K^{-}$              & $0.22^{+0.19}_{-0.12}\pm0.02(<0.58)$ & $4\pi^{+}3\pi^{-}$        & $0.31\pm0.08$ \\
                             &                                      & $4\pi^{+}3\pi^{-}\pi^{0}$ & $0.05\pm0.02$ \\
  \hline
  \hline
\end{tabular}
\end{center}
\end{table}

The sources of systematic uncertainties related to the measured cross sections are discussed in the following. The systematic uncertainties in the tracking and PID are $1\%$ per track~\cite{xiscatteringbes}. The uncertainty from the total number of $J/\psi$ events in data is reported in Ref.~\cite{totalnumber}. The uncertainties for the branching fractions are taken from the PDG~\cite{pdg}. To estimate the uncertainty from the position of the $\EE$ interaction point, we change the integral range by $\pm0.1$~cm~\cite{totalnumber}, which is from $(a, b)$ to $(a+0.1, b+0.1)$ or $(a-0.1, b-0.1)$, and the larger difference in the result is taken as the uncertainty. To estimate the uncertainty from the angular distribution of $J/\psi\too\Lambda\bar{\Lambda}$ and the $\bar{\Lambda}$ mean lifetime, we vary the angular distribution parameter $\alpha$ and the mean lifetime by $\pm1\sigma$. The uncertainty associated with the fit procedure arises from the background shape and fit range. The uncertainty associated with the background shape is estimated by changing a uniform distribution to a linear one, and the uncertainty from the fit range is obtained by varying the limit of the fit range by $\pm0.01$~GeV/$c^{2}$. The systematic uncertainties from the $\Lambda/\bar{\Lambda}/K_{S}^{0}$ mass windows, $R_{xy}$ requirement and $P(p_{\text{oil}})$ requirement are tested using a Barlow test method~\cite{lambdascatteringbes}, and are considered to be negligible. A summary of the systematic uncertainties is presented in Table~\ref{tab:sumerror}, and the total systematic uncertainty is obtained by adding all the individual components in quadrature assuming that they are independent. For all measured cross sections, statistical uncertainties are dominant, so the systematic uncertainties contribute little to the total uncertainties.

\begin{table*}[htbp]
\renewcommand{\arraystretch}{1.3}
\begin{center}
\caption{The summary of systematic uncertainties (in $\%$) in the cross section measurements. The symbol ``$-$" means that the source of systematic uncertainty is negligible.}
\label{tab:sumerror}
\begin{tabular}{lcccccc}
  \hline
  \hline
  Source \ \ & \ \ $K^{0}_{S}\pi^{+}$ \ \ & \ \ $K^{+}\pi^{+}\pi^{-}$ \ \ & \ \ $K^{0}_{S}2\pi^{+}\pi^{-}$ \ \ & \ \ $K^{+}2\pi^{+}2\pi^{-}$ \ \ & \ \ $K^{+}3\pi^{+}3\pi^{-}$ \ \ & \ \ $2K^{+}K^{-}$ \ \ \\
  \hline
  Tracking                                                 & 5.0 & 5.0 & 7.0 & 7.0 & 9.0 & 5.0 \\
  PID                                                      & 5.0 & 5.0 & 7.0 & 7.0 & 9.0 & 5.0 \\
  Number of $J/\psi$ events                                & 0.4 & 0.4 & 0.4 & 0.4 & 0.4 & 0.4 \\
  Branching fractions                                      & 4.3 & 4.3 & 4.3 & 4.3 & 4.3 & 4.3 \\
  $e^+e^-$ interaction point                               & 2.1 & 2.1 & 2.1 & 2.1 & 2.1 & 2.1 \\
  Angular distribution of $J/\psi\too\Lambda\bar{\Lambda}$ & 0.1 & 0.1 & 0.1 & 0.1 & 0.1 & 0.1 \\
  $\bar{\Lambda}$ mean life time                           & 0.2 & 0.2 & 0.2 & 0.2 & 0.2 & 0.2 \\
  Fit procedure                                            & 6.6 & 2.5 & 2.1 & 0.1 & $-$ & 0.1 \\
  \hline
  Sum                                                      &10.8 & 8.9 &11.2 &11.0 &13.6 & 8.6 \\
  \hline
  \hline
\end{tabular}
\end{center}
\end{table*}

\begin{figure}[htbp]
\begin{center}
\begin{overpic}[width=0.48\textwidth]{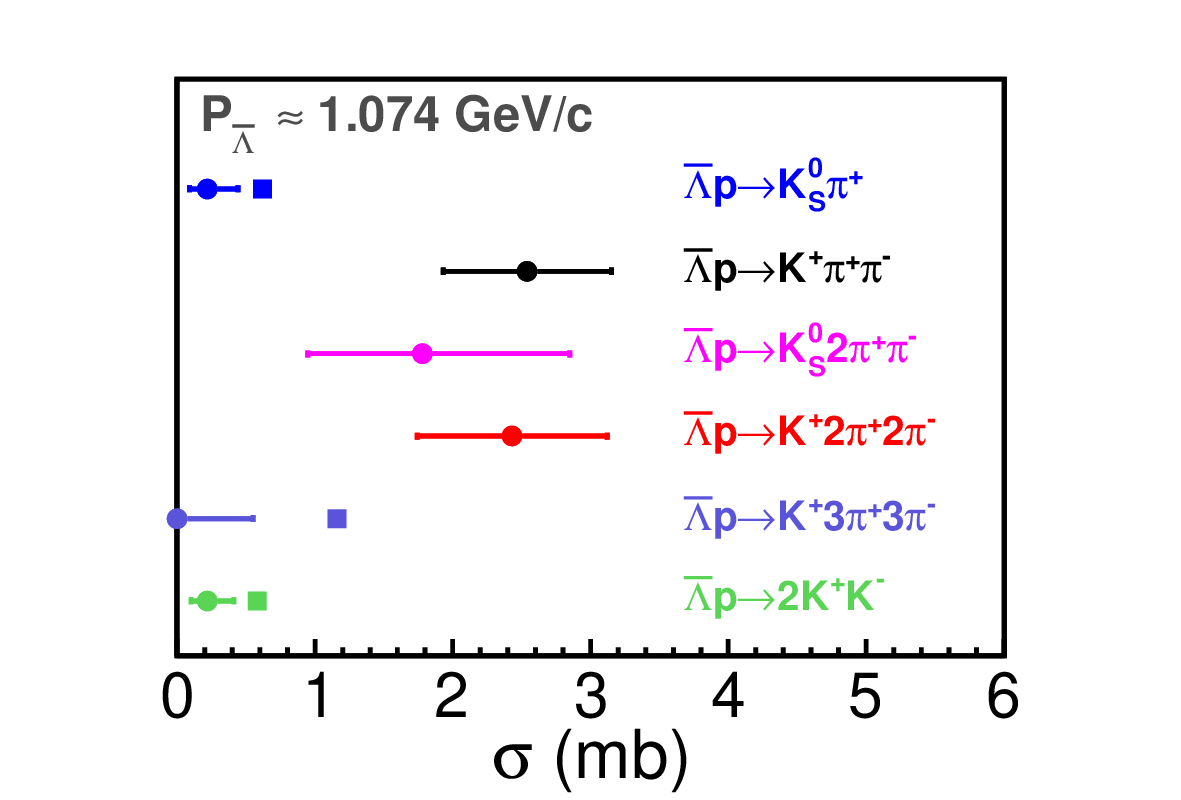}
\end{overpic}
\caption{Cross sections of the measured $\bar{\Lambda} p\too$ light mesons at the $\bar{\Lambda}$ momentum of around 1.074~GeV/$c$. The dots with error bars are the measured cross sections, which represents the total uncertainties, and the squares mark the upper limits at 90\% C.L. for the corresponding cross sections.}
\label{fig:fit_result2}
\end{center}
\end{figure}

In summary, using $(10.087\pm0.044)\times10^{9}$ $J/\psi$ events collected with the BESIII detector operating at the BEPCII storage ring, the annihilation of $\bar{\Lambda}$-$p$ into light mesons is studied, and the cross sections of these reactions are also measured at $P_{\bar{\Lambda}}\approx 1.074$ GeV/$c$. Among them, the reactions $\bar{\Lambda}p\too K^+\pi^+\pi^-$ and $K^+2\pi^+2\pi^-$ are observed, evidence for the reaction $\bar{\Lambda}p\too K^0_S2\pi^+\pi^-$ is found, while no significant signals are seen for the reactions $\bar{\Lambda}p\too K^0_S\pi^+$, $K^+3\pi^+3\pi^-$ and $2K^+K^-$. The results are listed in Table~\ref{tab:result2} and shown in Fig.~\ref{fig:fit_result2}. In addition, no significant structures are observed in the combinations of the final state particles for these reactions. The results of the annihilation of $\bar{\Lambda}$-$p$ and $\bar{n}$-$p$ into light mesons are compared together, and a similar behavior is found in this energy region. These new measurements will serve as important inputs for the theoretical understanding of the antihyperon-nucleon interaction, and further contribute to understanding antibaryon-baryon interactions in a unified perspective. Other antihyperon-nucleon annihilation into light mesons processes could also be studied in the future, such as $\bar{\Sigma} p\too$ light mesons and $\bar{\Xi} p\too$ light mesons. Such measurements would accelerate the experimental research progress in this field.

The BESIII Collaboration thanks the staff of BEPCII (https://cstr.cn/31109.02.BEPC) and the IHEP computing center for their strong support. This work is supported in part by National Key R\&D Program of China under Contracts Nos. 2023YFA1606000, 2023YFA1606704; National Natural Science Foundation of China (NSFC) under Contracts Nos. 12375071, 11635010, 11935015, 11935016, 11935018, 12025502, 12035009, 12035013, 12061131003, 12192260, 12192261, 12192262, 12192263, 12192264, 12192265, 12221005, 12225509, 12235017, 12342502, 12361141819; Natural Science Foundation of Henan under Contract No. 242300421163; the Chinese Academy of Sciences (CAS) Large-Scale Scientific Facility Program; the Strategic Priority Research Program of Chinese Academy of Sciences under Contract No. XDA0480600; CAS under Contract No. YSBR-101; 100 Talents Program of CAS; The Institute of Nuclear and Particle Physics (INPAC) and Shanghai Key Laboratory for Particle Physics and Cosmology; ERC under Contract No. 758462; German Research Foundation DFG under Contract No. FOR5327; Istituto Nazionale di Fisica Nucleare, Italy; Knut and Alice Wallenberg Foundation under Contracts Nos. 2021.0174, 2021.0299, 2023.0315; Ministry of Development of Turkey under Contract No. DPT2006K-120470; National Research Foundation of Korea under Contract No. NRF-2022R1A2C1092335; National Science and Technology fund of Mongolia; Polish National Science Centre under Contract No. 2024/53/B/ST2/00975; STFC (United Kingdom); Swedish Research Council under Contract No. 2019.04595; U. S. Department of Energy under Contract No. DE-FG02-05ER41374.

\end{document}